\documentclass[aps,pre,twocolumn,groupedaddress,showpacs]{revtex4}
\usepackage{amssymb}
\usepackage{graphicx,color}
\definecolor{r}{rgb}{1,0,0}
\definecolor{g}{rgb}{0,1,0}
\definecolor{b}{rgb}{0,0,1}
\begin{document}

%\preprint{}

\title{Morphology of rain water channelization in systematically varied model sandy soils }

% address or URL should go in the {}'s for \email and \homepage.

\author{Yuli Wei$^{1,2}$, Cesare M. Cejas$^{2}$, Remi Barrois$^{2}$, Remi Dreyfus$^{2}$, and Douglas J. Durian$^1$ }
%\thanks{}
\affiliation{$^{1}$Department of Physics and Astronomy, University of Pennsylvania, Philadelphia, PA 19104-6396, USA}
\affiliation{$^{2}$Complex Assemblies of Soft Matter, CNRS-Solvay-UPenn UMI 3254, Bristol, PA 19007-3624, USA}

%Collaboration name if desired (requires use of superscript address

\date{\today}

\begin{abstract}
We visualize the formation of fingered flow in dry model sandy soils under different raining conditions using a quasi-2d experimental set-up, and systematically determine the impact of soil grain diameter and surface wetting property on water channelization phenomenon.  The model sandy soils we use are random closely-packed glass beads with varied diameters and surface treatments.  For hydrophilic sandy soils, our experiments show that rain water infiltrates into a shallow top layer of soil and creates a horizontal water wetting front that grows downward homogeneously until instabilities occur to form fingered flows.  For hydrophobic sandy soils, in contrast, we observe that rain water ponds on the top of soil surface until the hydraulic pressure is strong enough to overcome the capillary repellency of soil and create narrow water channels that penetrate the soil packing.  Varying the raindrop impinging speed has little influence on water channel formation.  However, varying the rain rate causes significant changes in water infiltration depth, water channel width, and water channel separation.  At a fixed raining condition, we combine the effects of grain diameter and surface hydrophobicity into a single parameter and determine its influence on water infiltration depth, water channel width, and water channel separation.  We also demonstrate the efficiency of several soil water improvement methods that relate to rain water channelization phenomenon, including pre-wetting sandy soils at different level before rainfall, modifying soil surface flatness, and applying  superabsorbent hydrogel particles as soil modifiers.
\end{abstract}

\pacs{to be determined}
% list of pacs
%

%\keywords{}

\maketitle

%\section{}
%\subsection{}
%\subsubsection{}

% If in two-column mode, this environment will change to single-column
% format so that long equations can be displayed. Use
%\begin{widetext}
%\end{widetext}

%--------------------------------------------------------------------------------------------------

Improving the usage of rain and irrigation water by plants in sandy soils is an important topic in agriculture, which draws increasing attention with the reduction of water supply and the growth of the human population.  Sandy soils store water mainly through a capillary effect -- their pores capture and lock a small amount of water by capillary forces when rain or irrigation water flows through them.  Previous studies~\cite{Bhardwaj07, Bai10, Verneuil11, Wei13} have shown that superabsorbent hydrogel particle additives can significantly decrease the water conductivity and enhance water retention in sandy soils.  However, these studies were conducted in ideal fully-saturated soil systems, which significantly differ from a real situation in plant root zones commonly containing partially-wet or dry soils.  Early laboratory experiments~\cite{Hill72, Diment85, Glass89exp, Baker90} on water infiltration studies observed the formation of fingered flows in dry layered sands under uniform water flow onto the top sand layer.  Later, field studies~\cite{Vanommen89, Ritsema93, Hendrickx93, Ritsema98, Williams00} have determined the existence of preferential water paths in sandy soils during rainfall or irrigation.  At the same time, laboratory experiments~\cite{Selker92, Babel95, Yao96, Annaka10} have further confirmed that rain water channelization is a common feature that widely exists not only in sandy soils with structure heterogeneity, but also in uniform dry sands with almost no structure defects.  The cause of the latter one is due to instabilities that occur at the gravity-driven water wetting front~\cite{Selker92, Yao96}.  Rain water channelization largely reduces the water-reachable area in the plant root zone and results in a significant deviation of the predicted soil water capacity from measurements, which are usually started or performed at a fully-saturated state. Therefore, developing new techniques to incorporate water channelization phenomenon into the evaluation of soil water capacity and soil additive efficiency is crucial to achieve more reliable and applicable results.  To do so there is increasing need for characterizing the morphology of channelization, and understanding how it is affected by rain and soil properties.

This paper focuses on the {\it morphology} of rain water channelization.  After describing the experimental set-up, we examine rain water channelization using systematically varied dry model sandy soils with well-controlled grain diameters and surface wetting properties.  A quasi-2d set-up is built to mimic a steady rainfall and to capture the formation of water channels.  In steady state, the key parameters (including water infiltration depth, water channel width, and water channel separation) are determined for each soil sample and then plotted against the raining conditions (raindrop impinging speed and rain rate) or the soil properties (grain diameter and surface hydrophobicity).  Lastly, we discuss irrigation efficiency improvement methods that relate to rain water channelization phenomenon and demonstrate their effectiveness under different circumstances.  In a companion paper, we study the {\it kinetics} of rain water channelization \cite{Cejas2D}.

%---------------------------------------------------------------------------------------------------------------------

\section{Experiment}

For reproducible model sandy soils, we used mono-disperse solid glass beads with diameter varying from $D=0.18$~mm to $1$~mm (A-series, Potters Industries Inc.).  To clean the glass beads, they were first burned in a furnace at $500^{\circ}$C for $72$ hours and then soaked in a $1$M HCl bath for an hour.  After that the beads were rinsed with deionized water, baked in a vacuum oven at about $110^{\circ}$C for $12$ hours, and then cooled to room temperature in air.  The clean samples had hydrophilic surfaces -- our tests showed that the contact angle of water on a clean glass bead surface was $\theta^* = 16 \pm 2^{\circ}$.  Through additional chemical treatments, described below, we could modify the surface wetting property of clean glass beads to be hydrophobic -- our tests confirmed the contact angle of water on a treated glass bead surface was around $90^{\circ}$.  Mixing treated beads into the clean ones changes the effective contact angle of the whole packing, and the way we used to determine the effective contact angle of a soil packing (see Eq.~(\ref{ContactAngle})) is a set of independent capillary rise experiments rather than the contact angle measurement on a single glass bead.

We made two different size sample cells to hold the glass beads.  A $26$~cm wide and $30$~cm high sample cell was used when probing the effect of soil grain diameter; a $55$~cm wide and $15$~cm high sample cell was chosen when probing the effect of soil surface wetting properties.  All the sample cells were made of two parallels sheets of hydrophobic glass with a separation of $e=0.8$~cm.  The bottom of each sample cell was covered by several layers of meshes which held soil grains inside the sample cell well but allowed air to freely circulate in or out of the soil packing during rain.  Rain water could also freely drain out through the meshes as well.  The sample cell was cleaned and dried before each experiment.  Glass bead samples were then poured into the cell carefully.  During the pouring, we patted the cell gently from time to time to ensure a random close packing.  The volume fraction of glass beads is between $0.60$ to $0.64$.

\begin{figure}
\includegraphics[width=2.8in]{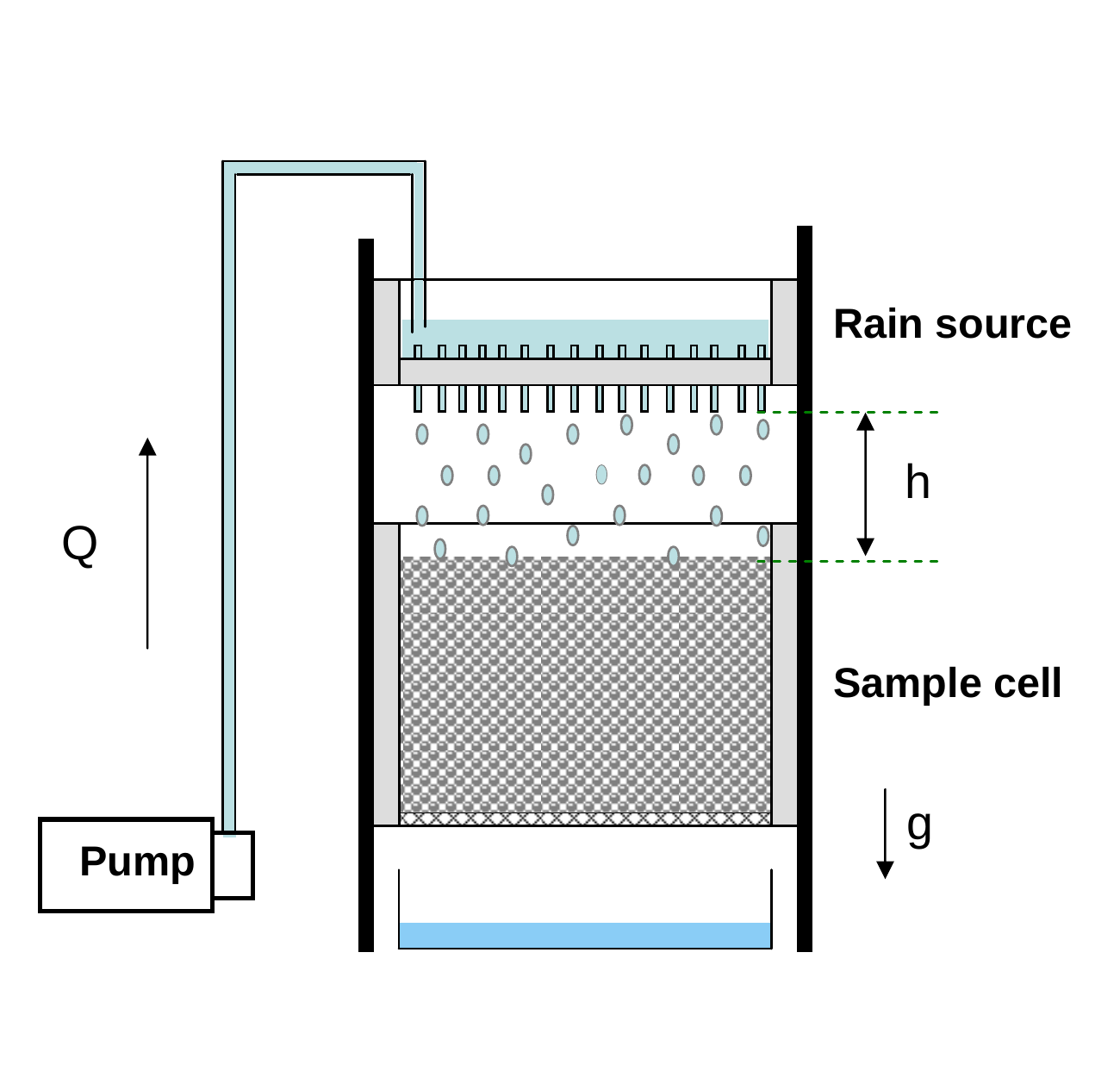}
\caption{(Color online) Schematic of the experimental set-up for visualizing rain water channelization in model sandy soils.  A sample cell partially filled with model sandy soils is suspended under a $2$D rain source built by a linear array of glass capillaries inserted into the bottom of a plastic container with a separation of $1$~cm between each other.  The sample cell can be shifted up or down to adjust the falling distance $h$ of rain droplets.  Rain rate $Q$ is determined by the water level in the plastic container.  A gear pump is used for supplying water during the experiment. }
 \label{Setup}
\end{figure}

Fig.~\ref{Setup} shows a schematic of the front view of our experimental set-up.  A sample cell was suspended under a 2d rain source built by inserting a line of glass capillaries ($5~\mu$L to $50~\mu$L Borosilicate Micro-Pipet, Kimble Inc.) in the bottom of a plastic container with a separation of $1$~cm between each other.  Rain rate $Q$ in units of cm/hr is defined as the volume of rain water per unit time per unit cross-sectional area of the sample.  The value was determined by measuring the mass of falling rain water within one minute for several times, and it showed a linear relationship with the water level in the plastic container.  A gear pump (Micropump Inc.) was used to maintain a constant water level in the plastic container.  The diameter of rain droplets was estimated by their average mass to be around $3$~mm.   The impinging speed $U_T$ of rain droplets depended on the free falling distance $h$ of rain droplets and was estimated as
\begin{equation}
   U_T = \sqrt{2gh}.
\label{Ut}
\end{equation}
Here, $g$ is the gravitational acceleration.  The location of the sample cell can be shifted up or down to adjust the value of $h$, and thus to vary the impinging speed $U_T$ of rain droplets.

During the experiments, we illuminated the sample cell from the back using a light box of the same size as the sample cell, and we take images from the front using a digital camera (Nikon D90).  The camera was controlled by a computer through a LabVIEW program which allowed us to automatically record an image sequence at a pre-set frame rate.

%---------------------------------------------------------------------------------------------------------------------------

\section{Water Channelization in Model Sandy Soils}

\begin{figure*}
\includegraphics[width=5in]{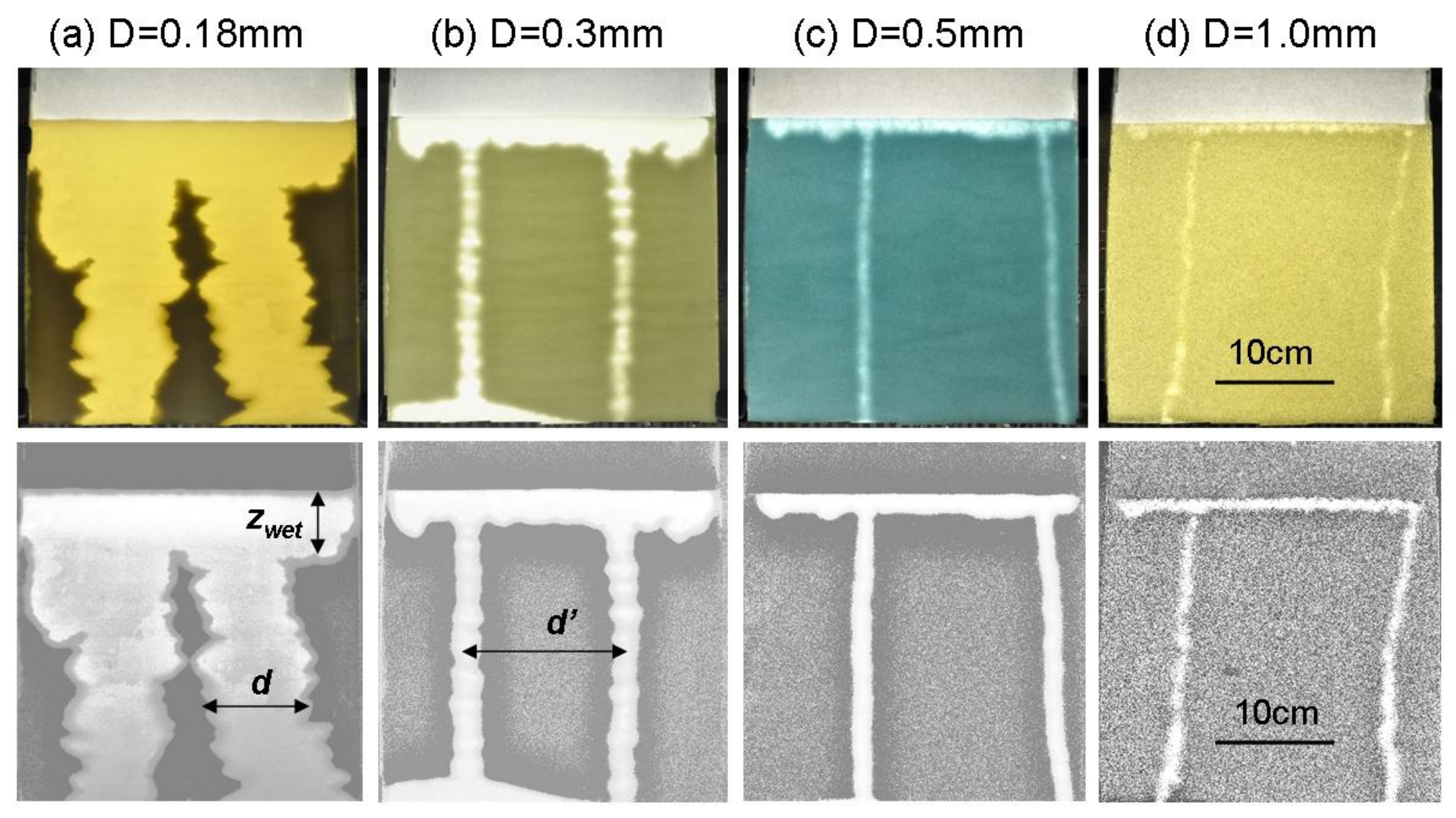}
\caption{(Color online)  Images (a) to (d) show deionized water channelization in dry hydrophilic model sandy soils with bead diameters $D$ varying from $0.18$~mm to $1$~mm, at a rain rate of $Q=14.5$~cm/hr and a raindrop impinging speed of $U_T=1$~m/s.  The color images in the first row are the original images taken at steady states; the grey-scale images in the second row are obtained by subtracting background images taken prior to rain from the ones taken at steady states, converting to grey-scale, and then enhancing contrast.  The sample packing in each image is $26$~cm wide, $25$~cm high, and $0.8$~cm thick.  As labeled in the images, $z_{wet}$ is infiltration depth of rain water in soils, $d$ is water channel width, and $d'$ is water channel separation.  The saturated hydraulic conductivity $\kappa_s$ (Eq.~(\ref{HydraulicConductivity})) for these four samples are determined as $73$~cm/hr, $204$~cm/hr, $567$~cm/hr, and $2300$~cm/hr, respectively.  }
\label{SizeEffect}
\end{figure*}

We begin with hydrophilic samples with varying grain diameters at different rain conditions.  Similar observations were obtained for these tests as described below and as demonstrated in Fig.~\ref{SizeEffect}.  Rain water infiltrates into a shallow top layer of soils to create a fully-saturated region with a horizontal wetting front.  As time goes on, the wetting front moves downward and the hydraulic pressure across it keeps increasing.  When it exceeds the capillary forces in model sandy soils, instabilities occur on the wetting front and grow to form water channels that penetrate through whole soil packing.  After that, rain water keeps flowing out of the soil packing through water channels and the system reaches a steady state.  We also noticed that during and after the formation of water channels, the wet regions (including the wet top layer and water channels) may become partially saturated.  The location of the water channels at different test runs varies but the separation between channels is very similar.  The use of different width sample cells excluded the effects of sample cell size.  As an example, in a $26$~cm wide packing of $1$~mm hydrophilic glass beads, we saw either one water channel formed near the center or two water channels formed close to each side; in a $55$~cm wide packing of $1$~mm hydrophilic glass beads, we saw $4$ to $6$ water channels formed with similar separation.

Fig.~\ref{SizeEffect}(a-d) show typical steady states of rain water channelization in dry hydrophilic model sandy soils with varying bead diameters.  The applied rainfall rate is $Q=14.5$~cm/hr, far smaller than the saturated hydraulic conductivity of soils.  The top row consists of raw images taken after steady state was achieved.  The bottom row shows grey-scale images obtained by subtracting background images taken before the rain started from the raw ones in the top row, converting the result to grey scale, and then enhancing the contrast using histogram equalization algorithm provided by MatLab.  The three parameters that we are interested in are labeled on the processed gray-scale images of the figure.  They are the infiltration depth $z_{wet}$, the channel width $d$, and the channel separation $d'$.  In the following subsections, we discuss the effects of raining conditions and sandy soil properties on these parameters respectively.

%---------------------------------------------------------------------------------------------------------------------------

\subsection{Effects of raindrop impinging speed and rain rate}

We first study the influence of raining conditions, such as raindrop impinging speed $U_T$ and rain rate $Q$, using $1$~mm dry hydrophilic glass bead packing.  Both deionized water and glycerol-water mixture are used as the rain water supply so that the effects of rain water quality can also be investigated.  An important parameter called the saturated hydraulic conductivity is widely used in these studies.  It quantitatively shows how fast a fluid can move through the pore spaces in a soil.  For the model sandy soils we used, the value of the saturated hydraulic conductivity $\kappa_s$ can be determined as
\begin{equation}
   \kappa_s = \frac{\rho g}{\mu}K_0 D^2  ~,
\label{HydraulicConductivity}
\end{equation}
where $\rho$ and $\mu$ are the density and the viscosity of the applied fluid respectively, $D$ is the soil grain diameter, and $K_0$ is the intrinsic permeability of the soil.  For a random close packing of spheres with a porosity of $\epsilon \approx 0.36$, $K_0$ has a value of $6.3\times 10^{-4}$ \cite{Beavers73, Verneuil11}.

\begin{figure}
\includegraphics[width=2.8in]{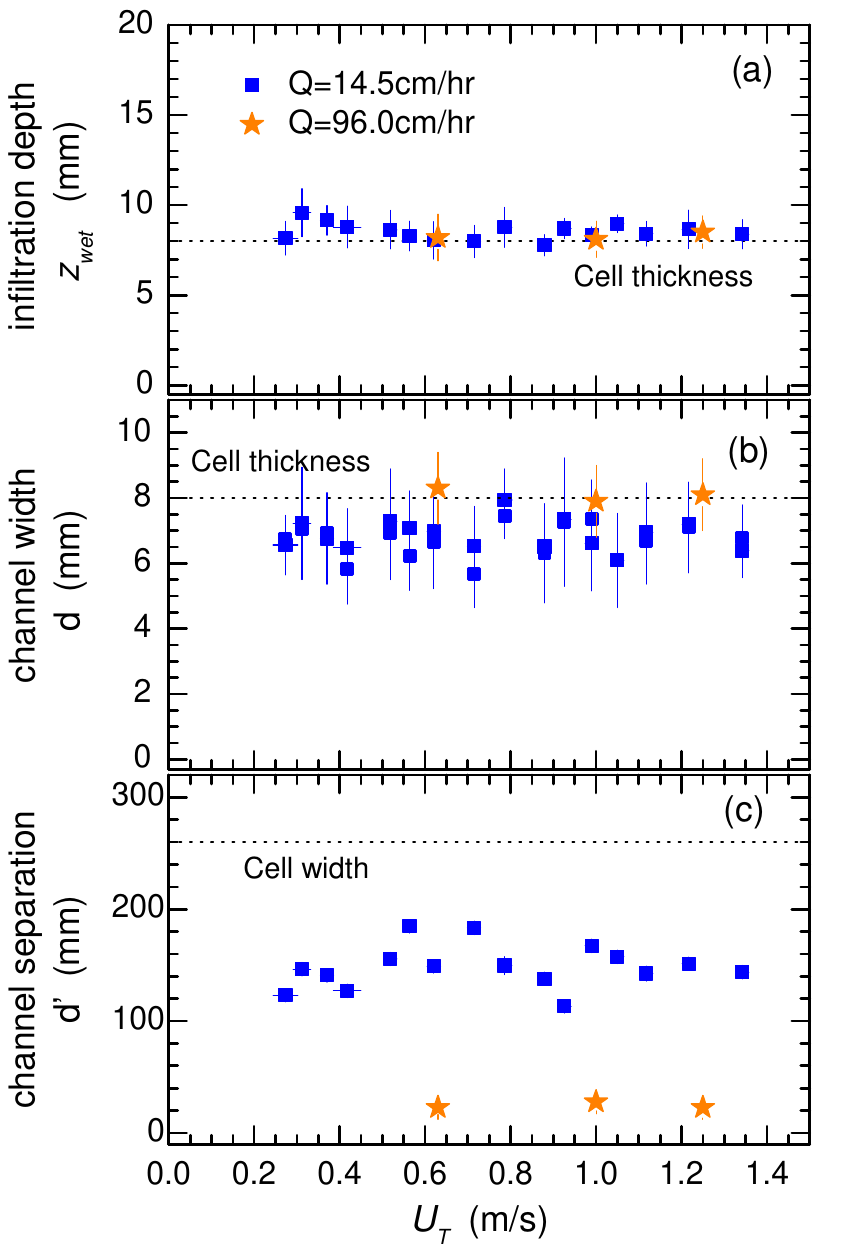}
\caption{(Color online)  Variation of infiltration depth ($z_{wet}$), channel width ($d$) and channel separation ($d'$), with raindrop impinging speed ($U_T$) in a dry model sandy soil of $D=1$~mm hydrophilic glass beads at two fixed rain rates.  The parameters are defined in Fig.~\ref{SizeEffect}.  Raindrop impinging speed is estimated from the falling height of the rain droplets using Eq.~(\ref{Ut}).  When a low rain rate of $Q=14.5$~cm/hr is applied, deionized water is used as `rain water' and the saturated hydraulic conductivity in the model sandy soil is determined to be $\kappa_s =2300$~cm/hr (Eq.~(\ref{HydraulicConductivity})).  When a high rain rate of $Q=96.0$~cm/hr is applied, a glycerol-water mixture is used as `rain water' and the corresponding saturated hydraulic conductivity in the same model sandy soil is reduced to be $\kappa_s =620$~cm/hr (Eq.~(\ref{HydraulicConductivity})).  }
 \label{VaryingUt}
\end{figure}

When impinging on sandy soils, rain droplets create craters on the soil surface, whose size and shape depend strongly on both the impinging speed and the soil grain diameter~\cite{Katsuragi10, Delon11}.  To determine if the craters affect the stability of the horizontal wetting front and thus control the locations of water channels, we varied raindrop impinging speed $U_T$ by adjusting the distance $h$ between the capillary tips on the rain source and the soil surface at two fixed rain rates, $Q=14.5$~cm/hr and $Q=96.0$~cm/hr.  When a low rain rate of $Q=14.5$~cm/hr is applied, deionized water is used as `rain water' and the saturated hydraulic conductivity in a model sandy soil is determined to be $\kappa_s =2300$~cm/hr using Eq.~(\ref{HydraulicConductivity}).  The ratio of rain rate over soil saturated hydraulic conductivity is only around $0.006$ for these cases.  Similar experimental observations were obtained for all $U_T$ values we have tested, as described before. Quantitatively, as shown in Fig.~\ref{VaryingUt}(a-c), there is no obvious difference in infiltration depth $z_{wet}$, channel width $d$, and channel separation $d'$, as the value of $U_T$ increases at a fixed low rain rate of $Q=14.5$~cm/hr.  The reason is that the infiltration depth in model sandy soils is larger than the size of the craters.  When a high rain rate of $Q=96.0$~cm/hr is applied, a $40$\% glycerol-water mixture~\cite{Cheng08} is used as `rain water' and the saturated hydraulic conductivity in the same model sandy soil is reduced to be $\kappa_s =620$~cm/hr due to the higher viscosity of the `rain water'.  The ratio of rain rate over soil saturated hydraulic conductivity now rises to be around $0.16$, but the experimental observation for different $U_T$ value cases is still roughly the same.  Fig.~\ref{VaryingUt}(a-c) again show that the raindrop impinging speed has no influence on infiltration depth $z_{wet}$, channel width $d$, and channel separation $d'$ at a high rain rate.  The reason is the same as that given for the low rain rate cases.

\begin{figure}
\includegraphics[width=2.8in]{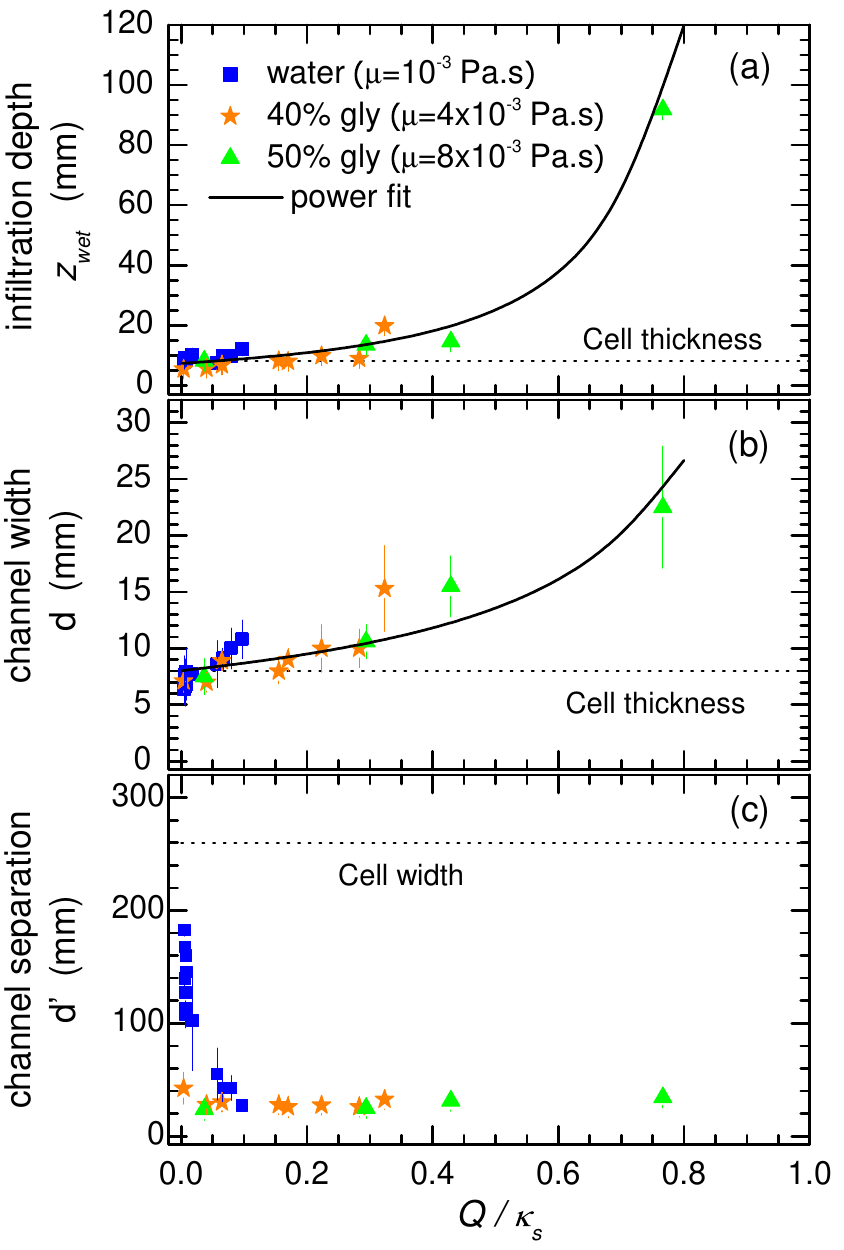}
\caption{(Color online)  Variation of infiltration depth ($z_{wet}$), channel width ($d$) and channel separation ($d'$), with rain rate ($Q$) in a dry model sandy soil of  $D=1$~mm hydrophilic glass beads, at a fixed raindrop impinging speed $U_T=1.0$~m/s.  The parameters are defined in Fig.~\ref{SizeEffect}.  Rain rate $Q$ is scaled by the saturated hydraulic conductivity $\kappa_s$ of the model sandy soil.  Three different liquids are used as `rain water' to extend the testing range: deionized water, a $40$\% glycerol-water mixture, and a $50$\% glycerol-water mixture.  Their corresponding saturated hydraulic conductivities in the same model sandy soil are determined to be $\kappa_s =2300$~cm/hr, $\kappa_s =620$~cm/hr and $\kappa_s =326$~cm/hr respectively (Eq.~(\ref{HydraulicConductivity})).  Solid lines in (a) and (b) are power-law fits to the data.   }
 \label{VaryingQk}
\end{figure}

Unlike raindrop impinging speed that has no influence at all, rain rate tells a very different story.  We notice that in Fig.~\ref{VaryingUt}(b) the channel width is slightly larger when a higher rain rate is applied and in Fig.~\ref{VaryingUt}(c) the channel separation is dramatically reduced at a higher rain rate.  To further clarify the influence of rain rate, we fixed raindrop impinging speed to be $U_T=1.0$~m/s and varied rain rate $Q$ by changing the water level in the water container.  Yao {\it et al}.~\cite{Yao96} have reported the unusual water channel size changes at extremely low rain rates using real sands with different grain size ranges.  Here we focus on a relatively high range of rain rates, from $12$~cm/hr to $220$~cm/hr, and apply different types of `rain water' in a model sandy soil with well-known pore structure.  Since many theory studies~\cite{Chuoke59, Parlange76, Glass89theory, Glass91, Cueto-Felgueroso08} suggested that the relation between the rain rate and the soil hydraulic conductivity played an important role in the formation and the size of the water channels, we scale the applied rain rate $Q$ with the saturated hydraulic conductivity of soil $\kappa_s$ in our plots to see if we could collapse the data obtained from different types of `rain water'.  As shown in Fig.~\ref{VaryingQk}, beside deionized water, a $40$\% glycerol-water mixture and a $50$\% glycerol-water mixture are applied in the experiments.  Their viscosity is about $4$ times and $8$ times higher than deionized water respectively, but their density and surface tension are still very close to deionized water (changes usually within $10$\%)~\cite{Cheng08, Shchekotov}.  Based on Eq.~(\ref{HydraulicConductivity}), using glycerol-water mixtures to replacing deionized water in the experiments reduces the hydraulic conductivity in the same model sandy soil to be $1/4$ and $1/8$ respectively and easily extend the range of $Q/\kappa_s$ to be close to $1$.

We observed water channel formation in all the experiments with varying channel position, size and number.  In Fig.~\ref{VaryingQk}(a), we see that the infiltration depths obtained from different testing liquids collapse in certain region.  As a whole, they show a strong dependence on the value of $Q/\kappa_s$ and rise dramatically as the value of $Q/\kappa_s$ approaches $1$.  It is reasonable since theoretically the stable horizontal wetting front should smoothly move downward forever when the supply water matches the saturated hydraulic conductively in a soil.  When the supply water is less than the soil saturated hydraulic conductivity, there is no need for all the soil to get wet to conduct the supply water -- thus instabilities may occurs on the wetting front and grow to be water channels.  We fit the data in Fig.~\ref{VaryingQk}(a) to a power law of $a (1-Q/\kappa_s)^{-\Delta}$ with $a$ and $\Delta$ as the fitting parameters, determining that $\Delta = 1.74\pm0.04$, as shown by the solid line in the figure.

The effects of varying $Q/\kappa_s$ on the water channel width is even more obvious.  In Fig.~\ref{VaryingQk}(b), the channel width data obtained from different viscosity liquids again collapse together and also show a power law increase as $Q/\kappa_s$ value increases.  We fit the data in Fig.~\ref{VaryingQk}(b) to the same power law equation of $b(1-Q/\kappa_s)^{-\Delta^*}$ with $b$ and $\Delta^*$ as the fitting parameters and obtain a value of $\Delta^* = 0.75\pm0.06$.

Previous theoretical studies by Chouke~\cite{Chuoke59} had shown that the water channel width followed this type of power law with a value of $\Delta^*$ equal to $1/2$.  Using a different analysis method, Parlange and Hill~\cite{Parlange76} later predicted that the water channel width followed the same power law but the value of $\Delta^*$ equals $1$.  Since both Chouke's model and Parlange's model aimed at real sandy soils, they used a measured hydraulic conductivity $\kappa_f$ of water channels rather than the saturated hydraulic conductivity $\kappa_s$ in their expressions (see Eq.~(\ref{Chouke}) and Eq.~(\ref{Parlange})).  For the model sandy soils we deal with, $\kappa_f \approx \kappa_s$ is a good assumption.  Comparing the values of $\Delta^*$, we find that our result lies coincidentally between Chouke's model and Parlange's model.  It is hard to say whose prediction may be better from here.  More detailed discussion regarding their models is given later in this paper when examining the effects of soil grain diameter and surface wetting property, and also in our kinetic study paper~\cite{Cejas2D}.

In Fig.~\ref{VaryingQk}(c), an interesting thing we notice is that the channel separation data obtained from deionized water at low $Q/\kappa_s$ values no longer collapse with those obtain from glycerol-water mixtures at similar $Q/\kappa_s$ values.  We believe the slight reduction of surface tension in glycerol-water mixtures compared to deionized water may contribute to this phenomenon.  When deionized water is applied in the experiments, we see that the number of channels formed in the sample cell increases quickly from $2$ to $9$ as rain rate $Q$ increases from $12$~cm/hr to $220$~cm/hr.  At the same time, the changes on the channel width is relatively small, only from about $7$~mm to $11$~mm.  The significant change on water channel number largely reduces the separation between channels, and thus we see a decrease in water channel separation in Fig.~\ref{VaryingQk}(c).  When glycerol-water mixtures are applied in the experiments, the number of channels formed in the sample cell varies between $6$ and $10$ as $Q$ increases from $2.3$~cm/hr to $250$~cm/hr.  But in the same $Q$ range their channel width is double or even triple.  Therefore, in Fig.~\ref{VaryingQk}(c), the channel separation obtained from glycerol-water mixtures are roughly the same at different $Q/\kappa_s$ values.

%---------------------------------------------------------------------------------------------------------------------------

\subsection{Effects of  grain diameter and surface wetting properties}

\begin{figure}
\includegraphics[width=2.8in]{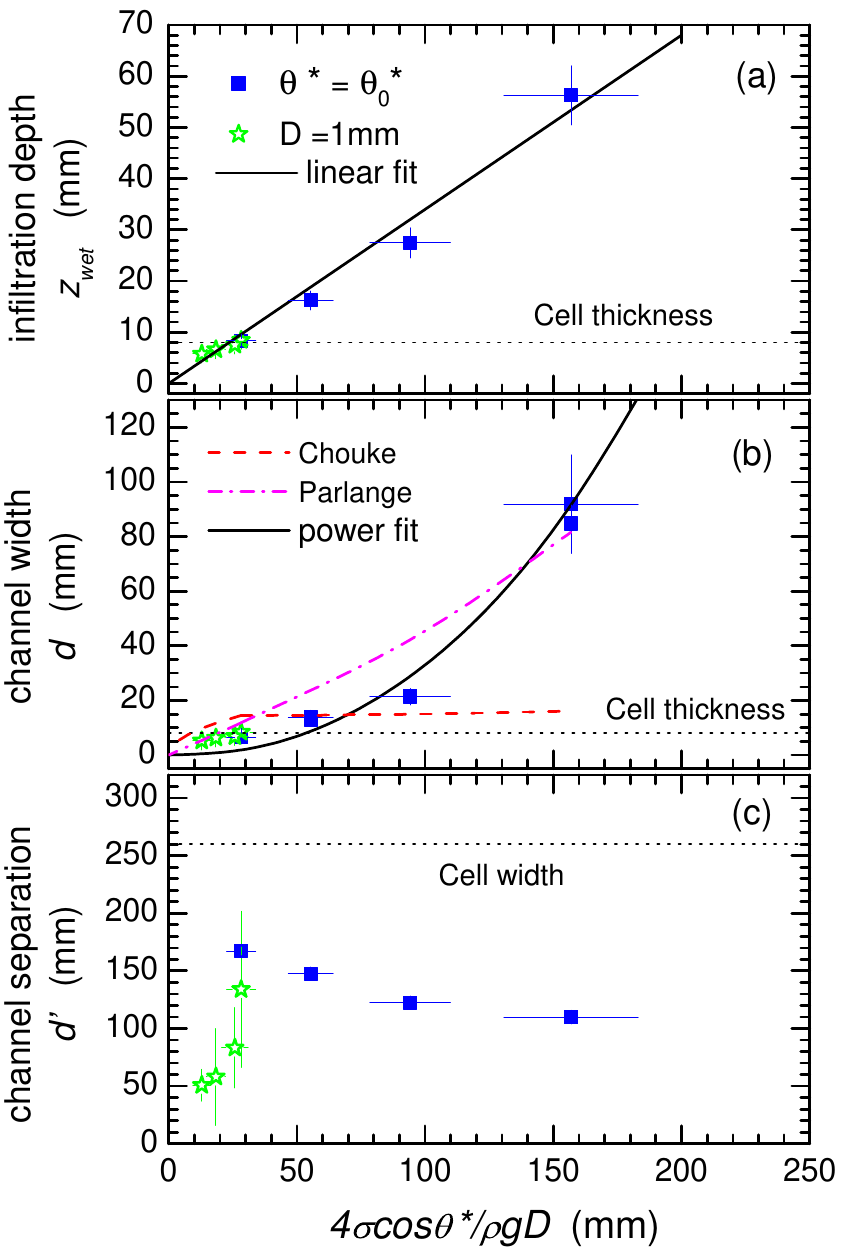}
\caption{(Color online)  Variation of the infiltration depth ($z_{wet}$), the water channel width ($d$), and the channel separation ($d'$), with the capillary rise height of the randomly close-packed glass beads ($4\sigma \cos\theta^*/\rho gD$) at a fixed raining condition.  The parameters plotted here have been defined in Fig.~\ref{SizeEffect}.  The rain rate is $Q=14.5$~cm/hr and the raindrop impinging speed is $U_T=1.0$~m/s.  In plot (a) to (c), solid squares represent the data obtained form hydrophilic glass beads with varied diameters (fix $\theta^*$ and vary $D$); while open stars represent data obtained from $1$~mm glass beads with varied surface wetting properties (fix $D$ and vary $\theta^*$).  Solid curve in (a) is a linearly fit to the data.  Solid curve in (b) is a power-law fit and the dashed curves are predictions from Eq.~(\ref{Chouke}) and Eq.~(\ref{Parlange2}) respectively.  }
 \label{VaryingD}
\end{figure}

In this subsection the raining conditions are fixed to be $Q=14.5$~cm/hr and $U_T=1.0$~m/s so that we can focus on the effects of varying the soil grain diameter or surface wetting properties.  Since the pore size of the sandy soils is proportional to their grain diameter, increasing the grain diameter $D$ enlarges the soil pores and lowers the capillary forces in sandy soils.  In Fig.~\ref{SizeEffect}(a-d), we see that varying the grain diameter $D$ significantly changes the infiltration depth of the rain water and the width of the water channel as well.  However, the number of water channels formed in the same size sample cell remains the same and the channel separation is similar to each other.  The quantitative results are plotted in Fig.~\ref{VaryingD} using solid squares.

\begin{figure}
\includegraphics[width=2.5in]{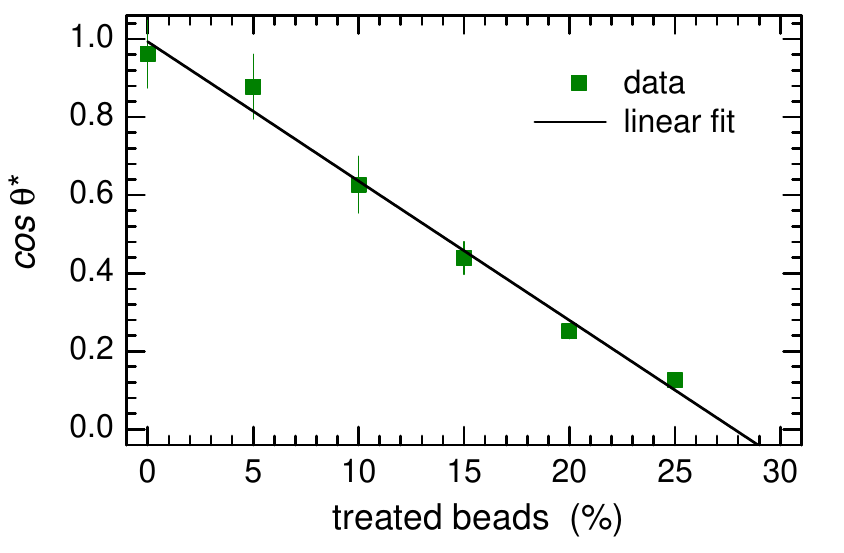}
\caption{(Color online) Variation of effective contact angle (cos$\theta^*$) with the percentage of hydrophobic beads.  The linear fit goes from complete wetting for untreated beads, to zero for a mixture of 30\% treated beads. }
\label{CapillaryRise}
\end{figure}

Besides the soil pore size, the surface wetting property also controls the capillary forces in sandy soils.  When the soil grains become partially hydrophobic, the capillary forces in soil pores drop quickly.  To determine the effects of soil hydrophobicity, we prepare partially hydrophobic samples by treating a small amount of clean $1$~mm glass beads with hydrophobic polymer solution (OMS Opto-chemicals, Montreal, Canada) and then uniformly mixing them into the same diameter untreated clean beads at different percentages.  Tests showed that the contact angle of water on a single treated bead was around $90^{\circ}$.  The hydrophobicity of the mixtures was determined by separate capillary rise experiments, similar to those employed by Durian {\it et al}.~\cite{Durian87}.  In capillary rise experiments, the air-dried mixtures were poured into hydrophobic glass tubes with meshes covered on their bottom.  Then the tubes were vertically placed in a shallow water reservoir for about $2$ days.  The difference of the water levels in glass bead packing and in reservoir gave us the value of capillary rise height $H$.   We measure that the contact angle of hydrophilic beads is $\theta^*_{0}= 16 \pm 2^{\circ}$. We also use its capillary rise height value $H_0$ as the reference to determine the effective contact angle for the mixture packing with both clean and treated beads:
\begin{equation}
 \cos\theta^* = \cos\theta^*_0 \frac{H}{H_0} ~.
\label{ContactAngle}
\end{equation}
From Eq.~(\ref{ContactAngle}), a glass bead packing of completely hydrophilic beads will have $\cos \theta^*\neq 1$ since the glass beads themselves are not perfectly hydrophilic as revealed from the measured contact angle value.  Furthermore, the capillary rise results for the mixtures are shown in Fig.~\ref{CapillaryRise}.  From the figure, we see that the value of cos$\theta^*$ decreases linearly as the percentage of treated beads increases.  When treated beads in the mixture reaches around $30$\%, the capillary rise height $H$ in the mixture decreases to zero and the mixture has an effective contact angle of $\theta^*=90^{\circ}$.

Under rainfall, the behavior of partially hydrophobic samples differs from that of very hydrophilic ones.  When the percentage of treated beads in a mixture is no more than $15$\% (cos$\theta^* =0.44$), we still see that rain water infiltrates into a top layer of soil samples and creates a horizonal wetting front.  After that, water channels form due to the instabilities and the system reaches steady state.  However, the infiltration depth in these cases is far shallower and the wetting front is not as flat and smooth as that seen in very hydrophilic case.  Also, a small change on soil hydrophobicity significantly increases the number of water channels that form in dry soils under rainfall.  For example, in a $55$~cm wide packing of glass beads, when treated beads change from $0$ to $15$\%, the number of water channels increases from $4$ to $10$.  For these cases, we try our best to extract their infiltration depth, channel width, and channel separation. The obtained values are also added to Fig.~\ref{VaryingD} in open stars.

When treated beads in a mixture exceeds $20$\% (cos$\theta^* =0.25$), rain water begins to pool on soil surface rather than infiltrates into soils.  As time goes on, the ponding water adds more and more hydraulic pressure on soil surface, which finally overcomes the soil water repellency and drives the formation of water channels.  Only one or two water channels were seen in these cases even in a $55$~cm wide sample cell.

Fig.~\ref{VaryingD} summarizes the results obtained from both hydrophilic samples (solid squares) and partially hydrophobic ones (open stars) at a fixed raining condition.  The expression in x-axis has a physical meaning of capillary rise height, which combines the influence of bead diameter $D$ and the effective contact angle $\theta^*$.  In Fig.~\ref{VaryingD}(a), we see that the infiltration depth in a soil sample grows almost linearly as the capillary rise height of that sample increases.  A linear fit on the data gives us a slope of $\alpha=0.34$.  Fig.~\ref{VaryingD}(b) shows the variation of water channel width.  As the capillary rise height of a soil increases, the water channel width increases quicker than the infiltration depth.  A power law fit is applied to the data and obtains a power value of $\delta =2.26$.  Fig.~\ref{VaryingD}(c) shows the variation of channel separation.  It remains more or less constant for hydrophilic samples with different bead diameters, but shows a clear decrease when soil sample become partially hydrophobic.

Among the parameters we discussed above, the water channel width is the only one that has been widely studied in literature~\cite{Chuoke59, Parlange76, Glass89exp, Glass91, Selker92, Yao96, Annaka10}.  Chouke {\it et al}.~\cite{Chuoke59} assumed a Laplace-Young relationship for the pressure jump across the horizonal wetting front and used a stability analysis to obtain an expression of the channel width as:
\begin{equation}
   d = \pi \left(\frac{3\sigma \cos\theta^*}{\rho g}\frac{1}{1-Q/\kappa_f }\right)^{1/2}  ~.
\label{Chouke}
\end{equation}
Here $\sigma$ is the surface tension of the applied rain water.  $\kappa_f$ is the measured hydraulic conductivity inside the water channel, whose value is usually a bit less or equal to the saturated hydraulic conductivity of the soil.  We assume $\kappa_f \approx \kappa_s$ and add the prediction of Eq.~(\ref{Chouke}) to Fig.~\ref{VaryingD}(b) in dashed line.  We see that Chouke's model significantly deviates from our experimental data, which is consistent with the conclusions Glass {\it et al}. drawn in their infiltration experiments on sands~\cite{Glass91}.

Parlange and Hill~\cite{Parlange76} later derived a relationship between wetting front velocity and curvature, and applied that in stability analysis to obtained an expression of
\begin{equation}
   d = \pi \frac{s_w^2}{\kappa_s(S_s-S_0)} \frac{1}{(1-Q/\kappa_f )}  ~.
\label{Parlange}
\end{equation}
Here $S$ is so-called water content and defined as the ratio of water volume retained in a soil and the total volume of the soil.  The subscripts `s' and `0' represent the saturated state and the initial state of the soil respectively.  For data shown in Fig.~\ref{VaryingD}(b), we have $S_0 =0$ (dry in initial state) and $S_s =\epsilon \approx 0.36$ ($\epsilon$ is the porosity of the sandy soil).

In Eq.~(\ref{Parlange}, $s_w$ is the soil water sorptivity, which has a unit of length over square root of time and measures the capacity of soils to absorb water through capillarity.  Culligan {\it et al}.~\cite{Culligan05} applied scaling analysis and showed that the soil water sorptivity can be calculated as
\begin{equation}
   s_w = s^*\left(\frac{\epsilon l^*}{\mu}\sigma \cos\theta^*\right)^{1/2}  ~.
\label{Sorptivity}
\end{equation}
Here $s^*$ is a dimensionless parameter called intrinsic sorptivity and its values is determined as $s^*=0.133$ by Parlange and his co-worker~\cite{Lockington03}.  $l^*$ is the characteristic pore radius in sandy soils and it is proportional to the diameter $D$ of soil grains.  We assume $l^*=\beta D$ and $\kappa_f \approx \kappa_s$.  Take Eq.~(\ref{Sorptivity}) into Eq.~(\ref{Parlange}), we have
\begin{equation}
   d = \beta \frac{\pi (s^*)^2 \sigma \cos\theta^*}{\rho g D K_0 }\frac{1}{(1-Q/\kappa_s )}  ~.
\label{Parlange2}
\end{equation}

We fit the data in Fig.~\ref{VaryingD}(b) using Eq.~(\ref{Parlange2}) and determine that $\beta=0.019$.  One may notice that the characteristic pore radius obtained here is far smaller that that determined by pressure plate method~\cite{Wei13}.  The reason is when writing down Eq.~(\ref{Sorptivity}) we ignore a term that reflect the saturation level in water channels and combine its influence into the fitting parameter $\beta$. Comparing the predictions from different models and a simple power-law fit, we find that the Parlange's model describes the experimental data better than the Chouke's model, which is consistent with previous literature~\cite{Glass89exp, Glass91, Selker92, Yao96}; however, the power-law fit offers an excellent empirical description.

%---------------------------------------------------------------------------------------------------------------------------

\section{Enhancing Rain Water Reachable Area in Sandy Soils}

\subsection{Pre-wetting sandy soils}

\begin{figure*}
\includegraphics[width=5.5in]{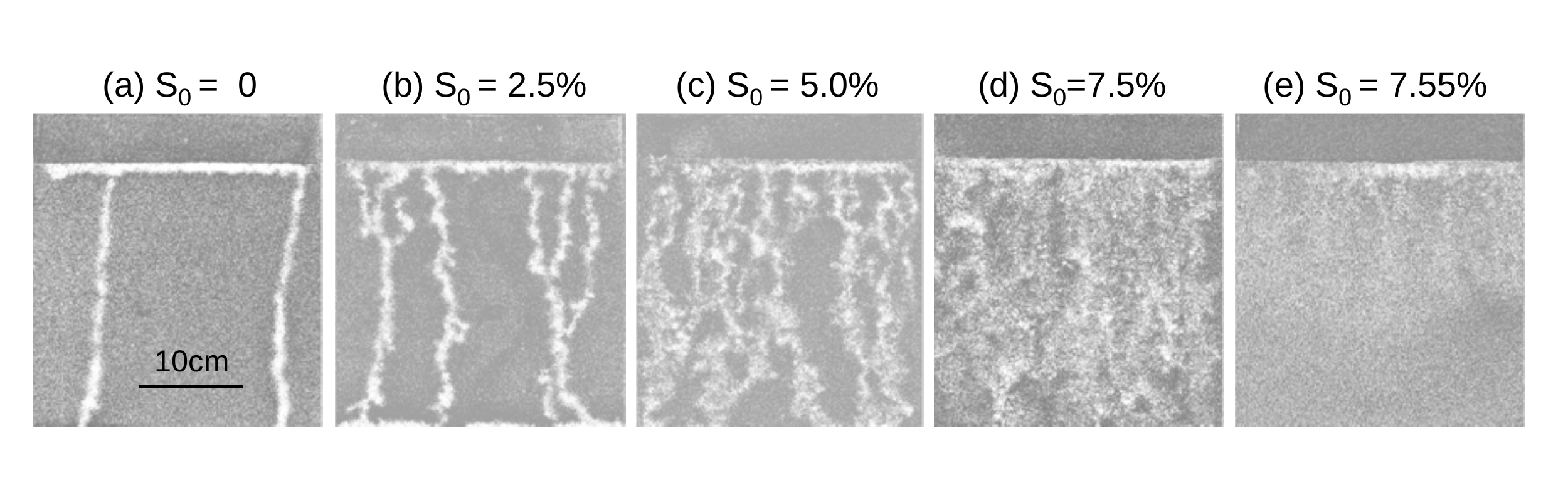}
\caption{ Grey-scale images showing the influence of initial water content ($S_0$) on rain water channelization in a model sandy soil, $1$~mm hydrophilic glass beads, at a fixed raining condition.  The rain rate is $Q=14.5$~cm/hr and the raindrop impinging speed is $U_T=1$~m/s.  The sample packing is $26$~cm wide and $0.8$~cm thick.  The sample packing height is around $25$~cm.  Image (a) to (e) show the steady state wetting patterns of rain water in a model sandy soil with initial water content $S_0$ increasing from $0$ to the filed capacity.  In this model sandy soil, the saturated water content is around $36$\% and the field capacity is around $7.75$\%.  }
 \label{PreWetting}
\end{figure*}

To improve the usage of rain and irrigation water by plants, an important strategy is to suppress or to modify rain water channelization behavior.  To suppress the water channelization behavior, a possible method is to pre-wet dry sandy soil a bit before rain.  Lu {\it et al}.~\cite{Lu94} observed that water channels initiated in a dry zone vanish in a pre-wet zone for a $2$D layered glass bead packing.  Later, Bauters {\it et al}.~\cite{Bauters00} determined that the width of the water channel formed in pre-wet quartz sand under a point water source shows strong dependence on the initial water content.  Here we systematically extracted the steady state wetting patterns of a pre-wet model sandy soil under non-ponding rainfall using $1$~mm hydrophilic glass beads.  For this, different amounts of water were uniformly mixed into air-dried glass beads and then the partially-wetted beads were closely packed into a sample cell.  The initial water content $S_0$ in each packing was determined as the ratio of the added water volume and the total volume of glass bead packing.  To estimate the so-called field capacity of the model sandy soil, which is the maximum amount of water a soil can hold in freely draining condition, we filled the sample cell with a fixed amount of air-dried model sandy soil, and then slowly immersed it into water to fully saturate the soil sample.  After that, the sample cell was moved out carefully and suspended in air for an hour of freely draining.  The wet soil sample was then poured out and the mass of the soil sample was measured.  By comparing the mass in wet and in dry, we determined that the field capacity of the model sandy soil was around $7.75$\%, which was only about one fifth of its full saturation value $36$\%.  The value of the field capacity set the upper limit of the initial water content we apply.

In the experiments, we set the rain rate to be $Q=14.5$~cm/hr and the raindrop impinging speed to be $U_T=1$~m/s.  The steady state wetting pattern at each initial water content value was obtained by subtracting the background image captured at rain start from that captured at the steady state and then by converting to grey-scale with enhanced contrast.  Fig.~\ref{PreWetting}(a-d) show the steady state wetting patterns for a model sandy soil with initial water content $S_0$ equal to $0$, $2.5$\%, $5.0$\%, and $7.5$\%, respectively.  We see more water channels forming when the initial water content $S_0$ increases.  Differing from the simulation results from Juanes's group~\cite{Cueto-Felgueroso08}, the uniform infiltration depth does not show a significant increase when the initial water content in soil increases.  Water channels formed in pre-wet soil have very irregular shapes, and sometimes they even branch or entangle together.  The equivalent channel width looks larger for soil with higher initial water content, which is consistent with the observation of Bauters {\it et al}.~\cite{Bauters00}.  For soil with an initial water content close to its field capacity, rain water is able to reach almost everywhere in the soil packing.  Fig.~\ref{PreWetting}(e) shows the steady state wetting pattern of rain water in a model soil packing with an initial water content equal to its field capacity.  This packing is prepared in the same way as we determined the soil field capacity.  It is used to exclude the possible structure heterogeneity due to the packing method.  We see that it is very similar to Fig.~\ref{PreWetting}(d).  Instead of forming water channels, the rain water flows throughout the soil packing.  Since pre-wet soil can effectively suppress rain water channelization and enlarge rain water reachable region in sandy soils, a good way to increase the usage of irrigation water by plants is to monitor the variation of the water content in soils with time and manage the irrigation time period based on that.

%---------------------------------------------------------------------------------------------------------------------------

\subsection{Modifying surface flatness of sandy soils}

\begin{figure*}
\includegraphics[width=4.5in]{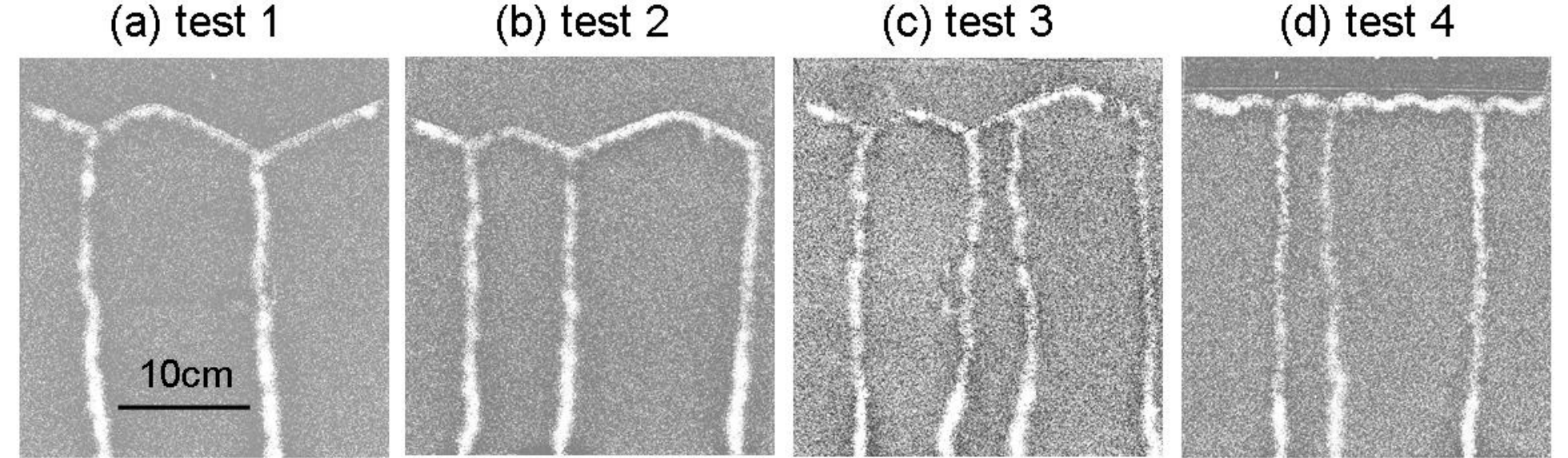}
\caption{ Grey-scale images showing the influence of surface flatness on rain water channelization in a dry model sandy soil, $1$~mm hydrophilic glass beads at a fixed raining condition.  The rain rate is $Q=14.5$~cm/hr and the raindrop impinging speed is $U_T=1$~m/s.  The sample packing is $26$~cm wide and $0.8$~cm thick.  The sample packing height varies due to the wave-like surface.  More water channels may form when wave-like surface creates an extra hydraulic pressure gradient on the wetting front.  }
 \label{WaveSurface}
\end{figure*}

Another method is to create an extra hydraulic pressure gradient on the horizontal wetting front to control the location and the separation of water channels.  By modifying the surface flatness of the soil packing, we are able to alter the shape of the wetting front away from horizontal.  Due to gravity, the hydraulic pressure on the bottom of the hill is higher than that at other locations.  Thus water channels should prefer to form there.  Fig.~\ref{WaveSurface} shows the steady state wetting pattern in a dry model sandy soil, hydrophilic $1$~mm glass beads, at four different test runs.  In each test, the raining conditions are set to be the same.  The soil surface shapes are set to be the same in test $2$ and test $3$; the soil surface shapes for other tests are different from each other.  We indeed find a curved wetting front in every test, such that the shape of the wetting front follows the shape of the soil surface.  In test $1$ and test $2$, water channels form at the bottoms of the hills, as expected.  But in test $3$, we see an extra channel forming on the slope of the hill.  And in test $4$, soil surface is curved to be a small wave with $7$ bottoms but only three channels finally grow.  These experiments show that we can increase the number of water channels in dry sandy soils by creating large enough curvatures on soil surface, but that we cannot precisely control their locations and numbers. 

%------------------------------------------------------------------------------------------------------------------

\subsection{Adding superabsorbent hydrogel particles}

As discussed above, by pre-wetting the soil or curving the soil surface, we can enlarge rain water-reachable region in sandy soils.  However, due to the high hydraulic conductivity in sandy soils, that amount of water will flow out quickly from the plant root zone unless soil additives are applied to help hold it in place.  Superabsorbent hydrogel particle additives are introduced for this purpose, and previous studies have proven their efficiency in reducing hydraulic conductivity in sandy soil and enhancing sandy soil water retention~\cite{Azzam83, Flannery82, Johnson84, Bouranis95, Bhardwaj07, Abedi08, Bai10, Wei13}.  The hydrogel particle additives we used here are a commercial product kindly provided by Degussa Inc.\ (Stockosorb SW).  The main chemical component of the product is cross-linked potassium acrylate and acrylamide at a ratio of $50/50$.  The dry hydrogel particles have a faceted shape and are sieved between $0.3$~mm and $0.5$~mm sized meshes.  When freely bathed with water, they swell to about $5$ times of their original diameter within about $3$ hours.  If the swollen hydrogel particles are then exposed in atmosphere at room temperature, we find that they de-swell slowly and lose over $80$\% of their stored water within a day.  These hydrogel particles are able to repeat the swelling to de-swelling cycle many times without decomposition.

Our previous $3$D raining experiments~\cite{Wei3D} have shown that these hydrogel particle additives significantly affect the transport and storage of rain water in sandy soils, and their effects strongly depend on their distribution methods.  Here, using the same $2$D set-up as above, we are now able to visualize the time evolution of hydrogel particle additives in sandy soil under rainfall.  Two different methods are applied to distribute dry hydrogel particles into a dry model sandy soil, $1$~mm hydrophilic glass beads.  One is uniformly mixing and the other is placing them in layer under ground.  Experiments are performed at a rain rate of $Q=14.5$~cm/hr and a raindrop impinging speed of $U_T=1$~m/s for over $5$ hours.  Dry hydrogel particles are white and too small to be seen when mixed into soil.  But after swelling they become large and transparent.  The back light passes through them to create bright spots, which can be easily identified in the raw images.

\begin{figure*}
\includegraphics[width=5.5in]{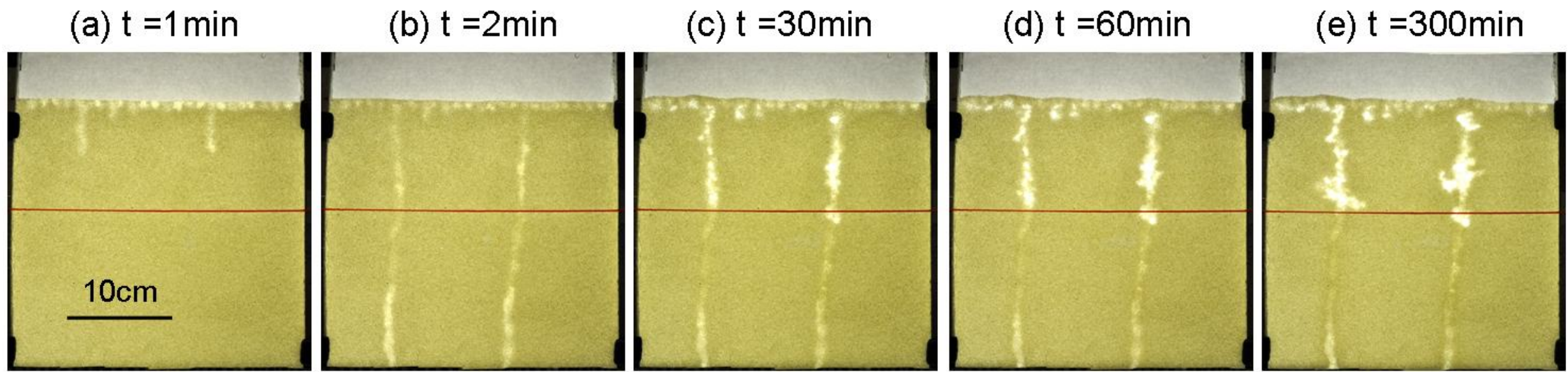}
\caption{(Color online)  Image sequence showing the influence of {\it uniformly-mixed} hydrogel particle additives on the rain water channelization in a sandy soil composed of $1$~mm hydrophilic glass beads.  $0.1$~wt\% dry hydrogel particles ($0.3$-$0.5$~mm in axis) are uniformly mixed into the top $10$~cm of the model sandy soil in dry before the rain.  The rain rate is $Q=14.5$~cm/hr and the raindrop impinging speed is $U_T=1$~m/s.  The sample packing is $26$~cm wide and $0.8$~cm thick.  The sample packing height is $25$~cm in dry.  The red line in each image indicates the boundary of the mixture and the pure model soil.  The swelling of hydrogel particles extends the water channels.  }
 \label{GelMixing}
\end{figure*}

\begin{figure*}
\includegraphics[width=5.5in]{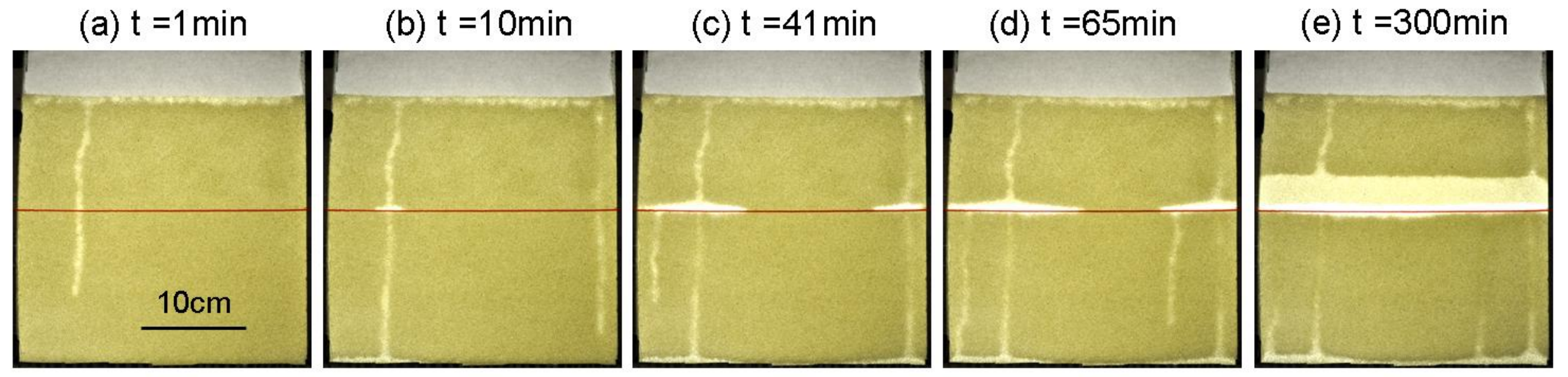}
\caption{(Color online)  Image sequence showing the influence of {\it a layer} of hydrogel particle additives on the rain water channelization in a sandy soil composed of $1$~mm hydrophilic glass beads.  $0.1$~g dry hydrogel particles ($0.3$-$0.5$~mm in axis) are placed in a layer under the top $10$~cm of the model sandy soil in dry before the rain.  The rain rate is $Q=14.5$~cm/hr and the raindrop impinging speed is $U_T=1$~m/s.  The sample packing is $26$~cm wide and $0.8$~cm thick.  The sample packing height is $25$~cm in dry.  The red line in each image indicates the location of the hydrogel particles.  The swollen hydrogel particles partially clog the water channels, form a wet hydrogel layer across the sample packing, and create a fully saturated region in soil above them.  }
\label{GelLayer}
\end{figure*}

Figs.~\ref{GelMixing}(a-e) show an image sequence captured from a $1$~mm hydrophilic glass bead packing with $0.1$~wt\% dry hydrogel particles uniformly mixed into the top $10$~cm region.  A red line is glued on the outside surface of the sample cell to mark the boundary of the mixture and the pure model soil.  We see that in this packing water channels form and grow the same as in a `no gel' packing at the early time of raining.  A bit later, hydrogel particles that are located at the top wet layer or in the water channels are swelling to a detectable size.  They begin to perturb the soil surface and extend the water channel above the red line.  As time goes on, the size of the swollen hydrogel particles keeps growing and the width of the water channels keeps enlarging.  Dry hydrogel particles nearby the water channel now contact rain water.  Finally, the water channels show very irregular shapes in the mixture region but keep the initial shape on the pure soil region.  Several conclusions are drawn from this experiment.  First, when mixed into soils, hydrogel particles enhance the soil water retention by extending the well-established water channels rather than by modifying the water channel formation process.  The reason is that the swelling time scale of the hydrogel particles is much slower than that of the water channel formation.  Second, within the range of concentration used here, the swollen hydrogel particles appearing in water channels cannot effectively clog them and reduce their hydraulic conductivity.  Third, rain water channelization keeps most of the hydrogel particle additives away from rain water, thus uniformly mixing a small percentage of dry hydrogel particles into sandy soils may significantly lower the efficiency of this product on improving soil water retention.

Fig.~\ref{GelLayer}(a-e) are an image sequence captured from a $1$~mm hydrophilic glass bead packing with $0.1$~gram dry hydrogel particles placed in a layer under the top $10$~cm of glass beads.  The red line shown in each image is glued on the outside of sample cell and it marks the location of the hydrogel particles in the model sandy soil packing.  We see that a water channel grows first after about $1$ minute of rain, and a second one grows about $10$ minutes later.  This situation happens commonly in pure model sandy soils.  The only difference is that a wet gel layer grows horizontally along the red line, where the dry hydrogel particles are placed.  As time goes on, the building wet gel layer begins to partially clog the water channels, thus two extra water channels form on the tips of the wet gel layer to drain out rain water.  When a continuous wet gel layer has formed, rain water begins to accumulate above it and a fully-saturated region grows upward in the model sandy soil.  Finally, the thickness of the wet gel layer and the depth of saturated soil region stop growing and a steady state is achieved.  In the steady state, the wet gel layer lifts the model sandy soil up a bit, all the added hydrogel particles are in swollen states, and some extra rain water is stored in the model sandy soil pores.  After rain stops, rain water stored in the model sandy soil keeps draining out slowly, but the water held by hydrogel particles stays for a long time.  This experiment clearly demonstrates that placing hydrogel particles in a layer under ground is a more efficient way to use this type of soil additive, especially when water channelization occurs in soils during rain or irrigation.

%------------------------------------------------------------------------------------------------------------------

\section{Conclusion}

In summary, we have investigated the phenomenon of rain water channelization in systematically varied model sandy soils that have well-established pore structure and surface wetting properties at different raining conditions.  In a homogenous sandy soil packing, the formation of water channels is caused by the instabilities occurring at the horizontal wetting front.  By visualizing water channels in sandy soils with different grain diameter and surface wetting properties at different raining conditions, we determine the relationships between the infiltration depth, the water channel width, the channel separation with the raining conditions and the capillary forces in model sandy soils.  Raindrop impinging speed has little influence on water channel formation, but rain rate strongly affects water channel size and separation.  The infiltration depth increases linearly as capillary forces increase; the water channel width increases even quicker than the infiltration depth, and is well-described by a power law; the channel separation shows almost no dependence on grain diameter but is very sensitive to even a small change on the soil wetting property.  Among these parameters, the water channel width is the only one that has been frequently discussed in literature.  We compare our results to former predictions on water channel width and found that our results obey the most commonly-used predictions from Parlange and Hill~\cite{Parlange76}.

With this understanding on rain water channelization in sandy soils, we then study the efficiency of different methods on improving rain water retention.  By pre-wetting a dry model sandy soil at different levels before rain, we can increase the number of water channels and enlarge the channel width, and thus extend the rain water reachable region.  By curving the soil surface to wave-like shapes, we can control the shape of the wetting front formed under rainfall, and thus create more water channels in dry sandy soils.  By adding superabsorbent hydrogel particles, we can modify the water channels in sandy soils in different ways by using different distribution methods.  Since the swelling time of hydrogel particles is far longer than that of the water channel formation, mixing a small amount into sandy soils has almost no effect on the formation of water channels in the early time of rain.  However, as time passes the swelling of hydrogel particles perturbs the soil structure and changes the wetting pattern in soils.  When uniformly mixed in soil, only the hydrogel particles that reside in the top wet layer or within the water channels are able to contact rain water.  Their swelling extends the water channels.  When placed in layer under ground, by contrast, all the hydrogel particles may be able to contact rain water sooner or later.  They not only swell to hold rain water inside, but also partially clog water channels and cause accumulation of rain water in the soil above.  Comparing the two methods, the latter one is a more efficient way to use hydrogel particle additives.

\begin{acknowledgments}
We thank Jean-Christophe Castaing, Larry Hough and Zhiyun Chen in Solvay-Rhodia Inc. for helpful discussions.  We also thank Christian Fretigny in CNRS for suggestions.  This work is supported by the National Science Foundation through grants MRSEC/DMR-1120901 (Y.W. and D.J.D) and DMR-1305199 (D.J.D.).
\end{acknowledgments}


\bibliography{Channeling_References}
%
%
% ****** End of file template.aps ******
\end{document}